\documentclass[doublecol]{epl2}

\usepackage{graphicx}
\usepackage{subfigure}
\usepackage{epstopdf}
\usepackage{grffile}
\usepackage{bm}
\usepackage{mhchem}
\usepackage{color}
\usepackage{hyperref}
\usepackage{amsmath}

\newcommand{\ket}[1]{{\vert #1\rangle}}

\newcommand{\braket}[2]{\langle#1\vert#2\rangle}
\newcommand{\eval}[3]{\langle#1\vert#2\vert#3\rangle}

\newcommand{\ud}{\mathrm{d}}

\newcommand{\IM}{\text{Im}}

\title{Double-pulse deexcitations in a one-dimensional strongly
  correlated system}

\author{Hantao Lu\inst{1,2} \and Janez Bon\v{c}a\inst{3,4} \and Takami
  Tohyama\inst{1}}
\institute{
  \inst{1} Yukawa Institute for Theoretical Physics, Kyoto
  University, Kyoto, 606-8502, Japan \\
  \inst{2} Center for Interdisciplinary Studies $\&$ Key Laboratory
  for Magnetism and Magnetic Materials of the MoE, Lanzhou University,
  Lanzhou 730000, China \\
  \inst{3} Faculty of Mathematics and Physics, University of
  Ljubljana, SI-1000 Ljubljana, Slovenia, EU \\
  \inst{4} J. Stefan Institute, SI-1000 Ljubljana, Slovenia, EU}


\abstract{
  We investigate the ultrafast optical response of the one-dimensional
  half-filled extended Hubbard model exposed to two successive laser
  pulses. By using the time-dependent Lanczos method, we find that
  following the first pulse, the excitation and deexcitation process
  between the ground state and excitonic states can be precisely
  controlled by the relative temporal displacement of the pulses. The
  underlying physics can be understood in terms of a modified Rabi
  model. Our simulations clearly demonstrate the controllability of
  ultrafast transition between excited and deexcited phases in
  strongly correlated electron systems. }

\begin{document}

\maketitle


Coherent manipulations of quantum states are essential prerequisites
for many advanced technological applications. One of the fundamental
phenomena in the field of nonlinear light-matter interactions are Rabi
oscillations describing the time evolution of a two-level quantum
system exposed to an external field. The Rabi oscillation and its
generalizations, together with related concept of quantum
interference, play central roles in the fields of coherent control and
manipulations. In decades, with the development of ultrafast optical
techniques, quantum interference experiments have advanced beyond
atomic/molecular systems into more complex solid state systems, such
as semiconductor heterostructures and quantum dots~\cite{Heberle:1995,
  Marie:1997, Stievater:2001, Kamada:2001il, Htoon:2002ee}.

Recently, there has been a steady increase in research and development
of materials with correlated electrons, due to their multiple novel
features that will potentially provide key ingredients for future
technological applications and innovations. For example, resulting
from the competing orders inherent in correlated
systems~\cite{Dagotto:2005ip}, it is plausible to obtain significant
response, and induce fast transitions from one phase to another with
relatively weak external pulse~\cite{Orenstein:2012bz}. In order to
achieve efficient high performance manipulations, it is imperative to
examine the ultrafast coherent dynamics of these systems when
subjected to external stimuli. In this Letter we show that under
specific conditions Rabi-like oscillations may be observed in a
prototype correlated system close to the charge density wave (CDW)
transition.

The one-dimensional (1D) Mott insulators, including \ce{Sr2CuO3} and
halogen-bridge Ni compounds~\cite{Yamashita:1999bt} as their
well-known representatives, have been considered as promising
candidates for future optoelectronic materials. Swift recoveries
(within picoseconds) to the insulating state in the photoinduced
insulator-to-metal transitions have been
noticed~\cite{Ogasawara:2000er, Iwai:2003cj}, which can be three
orders of magnitude faster than in
semiconductors~\cite{Nagai:2002}. Gigantic third-order optical
nonlinearity has also been reported~\cite{Okamoto:2000cv}. A good
theoretical starting point to study 1D Mott insulators is the extended
Hubbard model (see, {\em e.g.}, Ref.~\cite{Maekawa:2004} and
references therein). In the later, besides the on-site repulsion $U$,
nearest neighbor electron-electron interactions, denoted as $V$, are
also included. It is well known that when $V$ exceeds a critical
value, bound states of charge carriers, i.e., doublons
(double-occupied sites) and holes (empty sites), are
stabilized~\cite{Stephan:1996, Gebhard:1997, Shuai:1997}. In some
region of the interaction parameter space, the contribution to the
optical absorption spectrum from these bound states, known as
excitons, can give rise to an isolated $\delta$-like peak with energy
below the Mott gap. The influence of the low-lying excitonic
excitations on optical properties has been examined
theoretically~\cite{Jeckelmann:2003, Matsueda:2004ht} and
experimentally~\cite{Okamoto:1990}. Nevertheless, the role of the
exciton in the ultrafast optical dynamics has not been fully
addressed.

Using the time-dependent Lanczos method, we investigate the extended
Hubbard model at half filling. We are particularly interested in its
optical response to an external field composed of two consecutive
laser pulses, which resemble the pump-probe setup in spectroscopy
experiments. Following the time evolution of the system in the
ultrafast regime, we show that by carefully tuning the interval
between the two pulses, the system can be excited by the first pulse
and then deexcited by the second one. The conditions for observing
such on-and-off phenomenon and the underlying physics are
analyzed. More specifically, close to the CDW phase transition where
long-lived excitons represent optically active excitations, the
many-body system can be mapped onto an effective two-level system, and
accordingly, its dynamical response can be understood in terms of a
simple modified Rabi model. Alternatively, far from the CDW phase
transition, e.g., $V=0$, such on-and-off signal is diminished by the
continuity of its optical spectrum and the fast decay of charge
carriers into the continuum. This controllable, ultrafast switching
realized in this model, though conceptually simple, can be easily
generalized to other correlated systems as well, providing that
certain type of excitations can be selectively excited by carefully
tuned optical pumps.


The half-filled extended 1D Hubbard model is written as
\begin{eqnarray}
H&=&-t_h\sum_{i,\sigma}\left(c^{\dagger}_{i,\sigma}
  c_{i+1,\sigma}+\text{H.c.}\right)+U\sum_{i}\left(n_{i,\uparrow}-\frac{1}{2}\right)\nonumber
\\
&&\times\left(n_{i,\downarrow}-\frac{1}{2}\right)
  +V\sum_{i}\left(n_{i}-1\right)\left(n_{i+1}-1\right),
\label{eq:1}
\end{eqnarray} 
where $c^{\dagger}_{i,\sigma}$ ($c_{i,\sigma}$) is the creation
(annihilation) operator for an electron with spin $\sigma$ at site
$i$, $n_i=n_{i,\uparrow}+n_{i,\downarrow}$, $t_h$ is the hopping
constant, $U$ and $V$ are on-site and nearest-neighbor repulsion
strength, respectively. As $V$ increases, the charge carriers, i.e.,
doublons and holes, become energetically more favorable, since $V$
mediates an effective attraction between holes and doublons. It has
been shown that in large-$U$ limit, when $V$ exceeds a critical value
$\sim 2t_h$, excitonic bound states are formed in the photoexcitation
processes~\cite{Stephan:1996, Gebhard:1997}, while the signature of
these bound states emerges as a peak-like structure at the lower edge
of the optical spectrum, separated from the continuum. The strength of
the peak is enhanced with the increase of $V$~\cite{Jeckelmann:2003,
  Matsueda:2004ht}. At large $U$, a first-order phase transition from
the spin-density-wave (SDW) to the CDW phase takes place when
$V_c\approx U/2$~\cite{Nakamura:2000gk, Ejima:2007go}.

The external laser pulse is introduced via the Peierls substitution in
the hopping terms of the Hamiltonian~(\ref{eq:1}):
\begin{equation}
c^{\dagger}_{i,\sigma}c_{i+1,\sigma}\rightarrow
e^{iA(t)}c^{\dagger}_{i,\sigma}c_{i+1,\sigma},
\qquad
H\rightarrow H(t).
\label{eq:2}
\end{equation}
The time-dependent vector potential $A(t)$ is written in the
temporal gauge~\cite{Matsueda:2010ue, Lu:2012es, DeFilippis:2012jk}:
\begin{eqnarray}
A(t)&=&A_0 e^{-\left(t-t_1\right)^2/2t_d^2}\cos
\left[\omega_{\text{pump}}\left(t-t_1\right)\right]
\nonumber \\
&&+A_0e^{-\left(t-t_2\right)^2/2t_d^2}\cos
\left[\omega_{\text{pump}}\left(t-t_2\right)\right].
\label{eq:3}
\end{eqnarray}
Here, for simplicity, two pulses with identical shape are
assumed. Their temporal separation is given by $\Delta
t:=t_2-t_1$, the distance between the two peaks.  $A_0$ controls the
laser intensity, $t_d$ determines the widths of pulses, and
$\omega_{\text{pump}}$ is the pumping frequency.

In order to investigate the system's evolution under the influence of
the pulses, we employ the time-dependent Lanczos method, which is
originally described in Ref.~\cite{park:5870}, and has been developed
into one of the standard numerical methods in the study of
nonequilibrium dynamics of correlated
systems~\cite{Prelovsek:20111p}. The basic idea is that starting from
the Schr\"odinger equation $i\partial\psi(t)/\partial t=H(t)\psi(t)$
(we have set $\hbar\equiv 1$), the time evolution of the wave function
$\ket{\psi(t)}$ is approximated by a step-wise change of time $t$ into
small increments $\delta t$. At each step, in the Krylov subspace
constructed from $\ket{\psi(t)}$, the Lanczos basis with dimension $M$
is generated, resulting in the time evolution
\begin{equation}
\ket{\psi(t+\delta t)}\simeq e^{-iH(t)\delta t}\ket{\psi(t)}
\simeq\sum_{l=1}^M e^{-i\epsilon_l\delta t}\ket{\phi_l}\braket{\phi_l}{\psi(t)},
\label{eq:4}
\end{equation}
where $\epsilon_l$ and $\ket{\phi_l}$, respectively, are eigenvalues
and eigenvectors of the tridiagonal matrix using $M$ Lanczos
iterations. In the time-dependent Lanczos method, the approximation of
the finite-dimension cutoff $M$ is correct at least to the $M$-th
Taylor-expansion order in the time step $\delta
t$~\cite{Prelovsek:20111p}.

In the following discussions, we set $t_h$ and $1/t_h$ as energy and
time units, and we fix $U=10\,t_h$. Periodic boundary conditions are
imposed. The ground state in all cases represents the initial state,
i.e., $\ket{\psi(t=0)}=\ket{\text{GS}}$. The time increment $\delta
t=0.01$, and the Lanczos dimension cutoff is set to $M=30$.


We carry out numerical simulations for the Hamiltonian~(\ref{eq:2}) at
half filling. Here, we focus on the time dependence of the energy in
the resonance case, i.e., the pumping frequency $\omega_{\text{pump}}$
matches the positions of the absorption peaks in the optical
spectra. The time-dependent energy is defined as
\begin{equation}
E(t)=\eval{\psi(t)}{H(t)}{\psi(t)}-E_{\text{GS}}.
\label{eq:5}
\end{equation}
Note that $E(t)$ can be also equally obtained by the integral
$-\int_0^{t}j(\tau)(\ud A(\tau)/\ud\tau)\ud\tau$, which measures the
absorbed energy. $j(t)$ is the expectation value of the field-induced
current operator:
$\hat{j}(t):=it_h\sum_{i,\sigma}[e^{iA(t)}c^{\dagger}_{i,\sigma}c_{i+1,\sigma}-\text{H.c.}]$,
and $j(t)=\eval{\psi(t)}{\hat{j}(t)}{\psi(t)}$.

\begin{figure}
\includegraphics[width=0.4\textwidth]{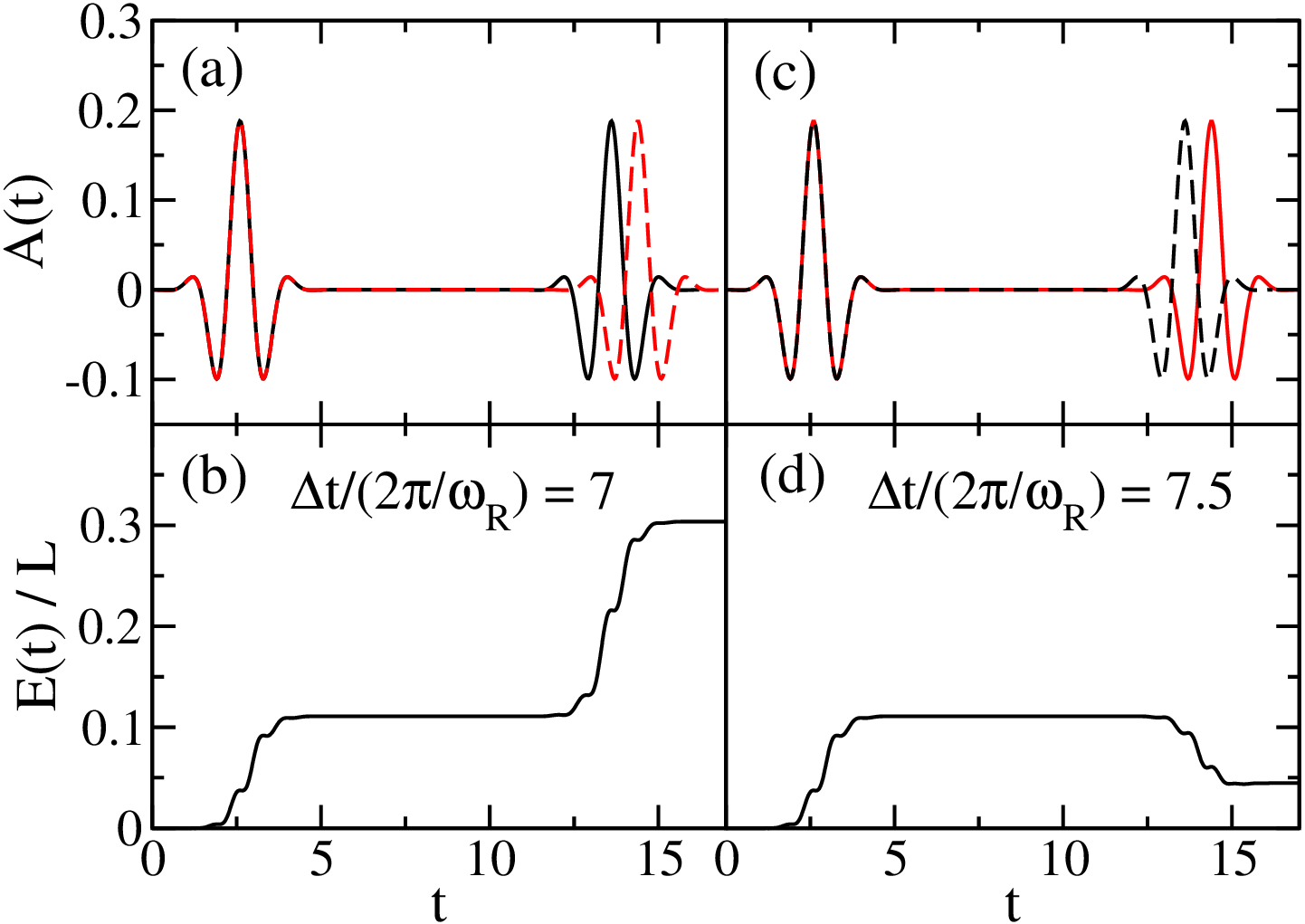}
\caption{(Color online) Time dependence of the vector potential $A(t)$
  ((a) and (c)) and the energy (measured from the ground state) per
  site $E(t)/L$ ($L$ is the lattice size) under the action of two
  consecutive pulses ((b) and (d)), for 14-site lattice with $U=10$,
  $V=4.5$. Parameters for the laser pumps: $A_0=0.19$,
  $\omega_{\text{pump}}=\omega_{\text{R}}=4.0$, $t_d=0.65$. The first
  peak of the pulses is fixed at $t_1=4\,t_d$; the second peak is
  determined by $t_2=t_1+\Delta t$. $\Delta t=7$ ((a) and (b)) and
  $7.5$ ((c) and (d)), respectively, in the unit of
  $2\pi/\omega_{\text{R}}$. In the upper part, the solid lines
  represent the shapes of the corresponding vector potential $A(t)$,
  while the other case is shown as a dashed line for comparison. The
  results of $E(t)$ are shown in the lower part.}
\label{fig:1}
\end{figure}

Our main results are shown in Fig.~\ref{fig:1} for 14 lattice sites at
$V=4.5$, where the ground state is in the SDW phase. The exciton state
is an excited state with the energy $\omega_{\text{R}}=4.0$ above the
ground state for this particular system size, and we put
$\omega_{\text{pump}}=\omega_{\text{R}}$. We deliberately chose a
small value of $t_d=0.65$, while the reason will be given later. The
shape of $A(t)$ is shown in Figs.~\ref{fig:1}(a) and (c). Suppose that
$t_h=1\ \text{eV}$, then in the real time unit, $\hbar/t_h=0.66\
\text{fs}$, $t_d=0.43\ \text{fs}$. The intensity of the laser pulse
$A_0=0.19$ is used so that the number of photons (per site) during the
pulse, $N_{\text{pt}}\propto A_0^2\omega_{\text{pump}}t_d$, is around
$0.5$~\cite{DeFilippis:2012jk}, which means that we confine our
calculations in the weak excitation regime~\cite{Htoon:2002ee}. The
maximum of the electric field can reach around $7.6\times 10^{9}\
\text{V/m}$ if the lattice constant is $1\ \text{\AA}$.

The main message in Fig.~\ref{fig:1} is that by tuning $\Delta t$, the
final energy $E_{\text{fin}}$ after the second pulse can be either
enhanced (Fig.~\ref{fig:1}(b)) or reduced (Fig.~\ref{fig:1}(d)),
compared with the intermediate stage energy (the energy between the
two pulses, which can be estimated from the plateau of $E(t)$ in
Fig.~\ref{fig:1}). It turns out that the beat is controlled by the
energy difference between the ground state and the excited states,
i.e., $\omega_{\text{R}}$. As shown in Fig.~\ref{fig:1}, we notice
that if $\Delta t=N\times\frac{2\pi}{\omega_{\text{R}}}$, where
$N\in\text{Integer}$, the highest enhancement is found; on the other
hand, the strongest suppression appears when $\Delta
t=\left(N+\frac{1}{2}\right)\times\frac{2\pi}{\omega_{\text{R}}}$. It
is noted that by tuning parameters, we are able to obtain almost
perfect energy suppression after the second pulse. For example, for
$t_d=5$, $A_0=0.01$, the corresponding $N_{\text{pt}}\sim 0.01$,
$E_{\text{fin}}$ turns out to be $0.003$.

The underlying physical picture can be captured by the Rabi
model~\cite{Rabi:1936, Rabi:1937}. The latter describes a two-level
quantum system coupled with a single-mode external field, and can be
written as
\begin{equation}
H_{\text{R}}(t)=\epsilon\sigma_z+g(t)\sigma_x,
\label{eq:7}
\end{equation}
where $\sigma_a$ are Pauli matrices, $2\epsilon$ is the level spacing;
the off-diagonal term $g(t)$, which describes the coupling of the
two-level system to an external field, typically has the form
$g(t)=2g\cos\omega t$. When $\omega\approx 2\epsilon$, the
rotating-wave approximation can be applied~\cite{Jaynes:1963fa}.  The
system's oscillation between the two levels, known as Rabi
oscillation, is fully characterized by two quantities, $\delta$ and
$\Omega_R$, known as detuning and Rabi frequencies, respectively:
\begin{equation}
\delta=\epsilon-\omega/2,
\qquad
\Omega_{R}=\left[(\epsilon-\omega/2)^2+g^2\right]^{1/2}.
\label{eq:10}
\end{equation}
At the tuning point, i.e., $\omega=2\epsilon$, only $\Omega_{R}$
remains, and is solely determined by the coupling strength $g$.

\begin{figure}
\includegraphics[width=0.45\textwidth]{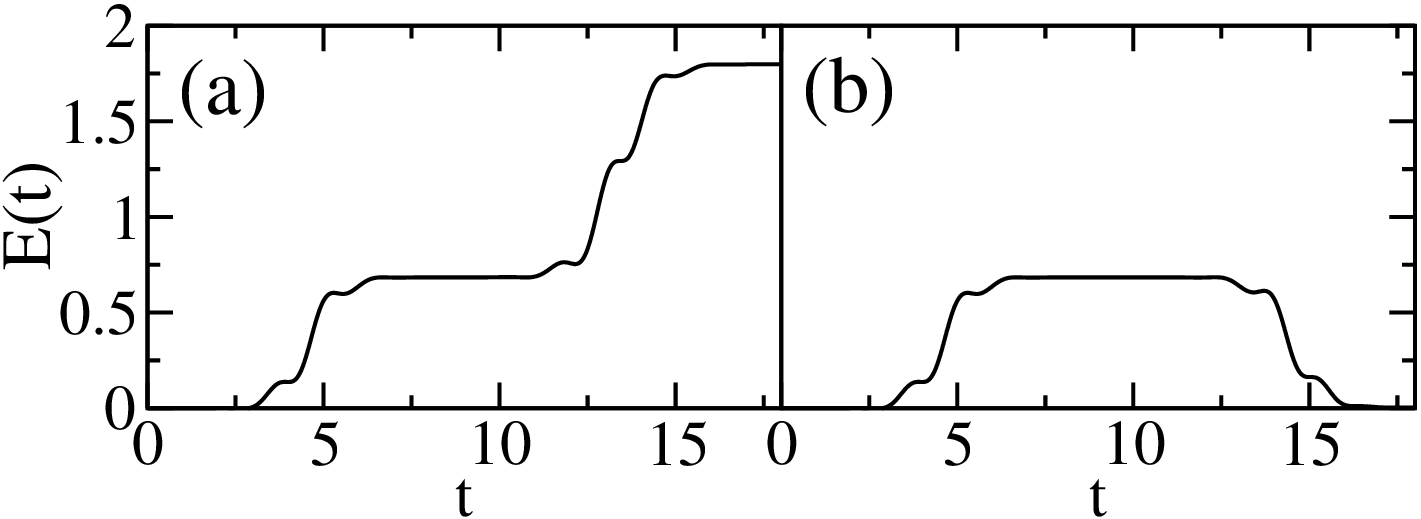}
\caption{The time-dependent energy expectation $E(t)$ (measured from
  the lower level) for the Rabi model coupled with two successive
  pulses. Here, we take $\epsilon=1$. Parameters for $A(t)$:
  $A_0=0.5$, $\omega_{\text{pump}}=2\epsilon=2$, $t_d=1$,
  $t_1=4t_d$. In (a), $\Delta t=3\times\pi/\epsilon$; in (b) $\Delta
  t=3.5\times\pi/\epsilon$.}
\label{fig:2}
\end{figure}

Keeping this picture in mind, we extend this model to a case when
$g(t)$ in Eq.~(\ref{eq:7}) is substituted by $A(t)$
(Eq.~(\ref{eq:3})). Figure~\ref{fig:2} shows the energy expectation of
the modified Rabi model at the tuning frequency
$\omega_{\text{pump}}=2\epsilon$. A typical quantum interference
phenomenon is observed. We can see that after a coherent exciton
polarization is created by the first pulse, the system energy $E(t)$
is either raised (Fig.~\ref{fig:2}(a)) or reduced
(Fig.~\ref{fig:2}(b)) by the second pulse, depending on the relative
phase of the two pulses~\cite{Kamada:2001il}.

\begin{figure}
\includegraphics[width=0.45\textwidth]{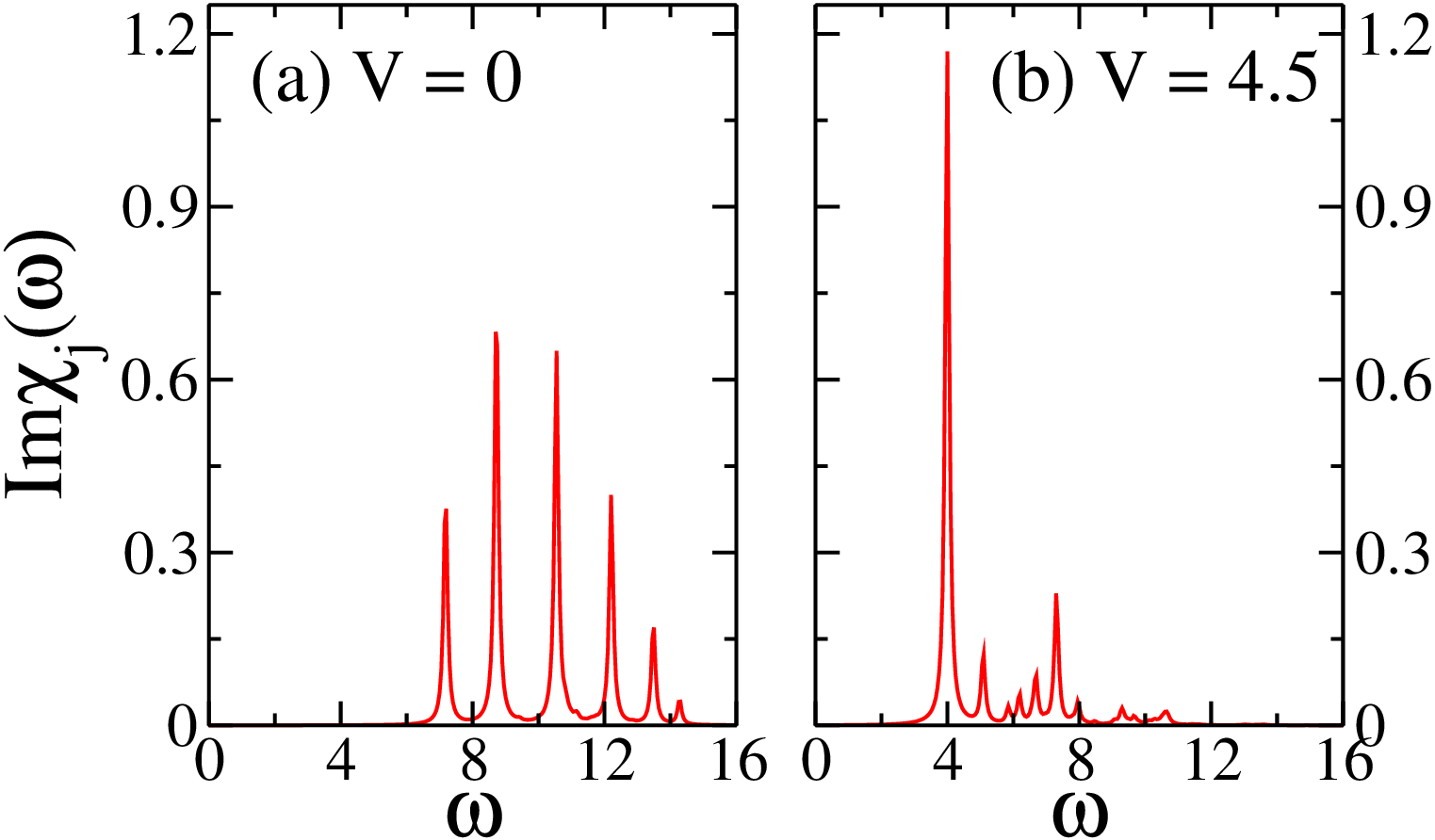}
\caption{The imaginary part of the dynamical current-current
  correlation function $\IM\chi_j(\omega)$ for $L=14$ lattice size with
  $V=0$ and $4.5$, obtained by Lanczos
  method~\cite{Lu:2012es}.}
\label{fig:3}
\end{figure}

The plain similarity between Figs.~\ref{fig:1} and \ref{fig:2}
suggests that the two-level model can provide qualitative
understanding of the optical response of the extended Hubbard
model. The analogy is justified by the existence of excitonic peak in
the optical absorption spectrum~\cite{Matsueda:2004ht} for the given
value of $V=4.5$, which is sharp and isolated from the continuum. (For
instance, see Fig.~\ref{fig:3}(b). Here in Fig.~\ref{fig:3}, we
present the optical absorption spectrum on $L=14$ lattice size for
$V=0$ and $4.5$. Note that the absorption spectrum is proportional to
the imaginary part of the current-current correlation function defined
by $\chi_j(\omega)=\eval{0}{\hat{j}[\omega-H+E_0]^{-1}\hat{j}}{0}$,
where $H$ is the time-independent Hamiltonian (Eq.~(\ref{eq:1})),
$\hat{j}$ is the current operator, $E_0$ the ground state energy.) For
this reason the decay time of the excitonic state is much slower than
$\Delta t$ which enables nearly complete deexcitation after the second
external pulse. The validity of this mapping can be justified
numerically based on the small lattice size simulations, where the
oscillation between the ground state and the specific excitonic state
extracted from the time evolved wave function $\ket{\psi(t)}$ has been
observed. It should be noted that the study on coherent dynamics
between several inherent energy levels under the influence of the
external stimuli is a widespread practice as mentioned in the
introduction. Yet to our knowledge, quantitative investigations
concerning the possibility of this kind of setup on correlated systems
are few and far between.

Naturally, we may wonder what will happen in the case of $V=0$, where
instead of well-defined excitonic peak, a continuum absorption is
expected in the thermodynamic limit, reflecting the response from
unbound charge carriers. In order to answer this question, we perform
the calculation on various lattice sizes, $L=10,\,12$ and $14$. Due to
the finite-size effects, a series of separate peaks with comparable
heights, representing a continuum in the thermodynamic limit, appear
in the optical absorption spectrum, e.g., see Fig.~\ref{fig:3}(a) for
$V=0$ on $L=14$ lattice size. The number of large peaks scales
linearly with the system size $L$, which can be interpreted as the
finite-size precursor of the continuum. Based on this observation, our
strategy is the following: we set the center of pumping frequency
$\omega_{\text{pump}}$ to match the first major peak in the optical
spectrum, denoted as $\omega_1$, while $t_d$ is determined by the
difference between the first and second resonance frequencies, i.e.,
$t_d=1/\left(\omega_2-\omega_1\right)$. This is the same approach as
we have already used for the calculations of Fig.~\ref{fig:1}. In this
way, the multi-frequency absorption, which is expected in the
thermodynamic limit when $V=0$, can be approximately simulated. For
various lattice sizes, we use identical $N_{\text{pt}}\sim 0.5$. The
laser intensity $A_0$ can then be determined accordingly.

\begin{figure}
\includegraphics[width=0.45\textwidth]{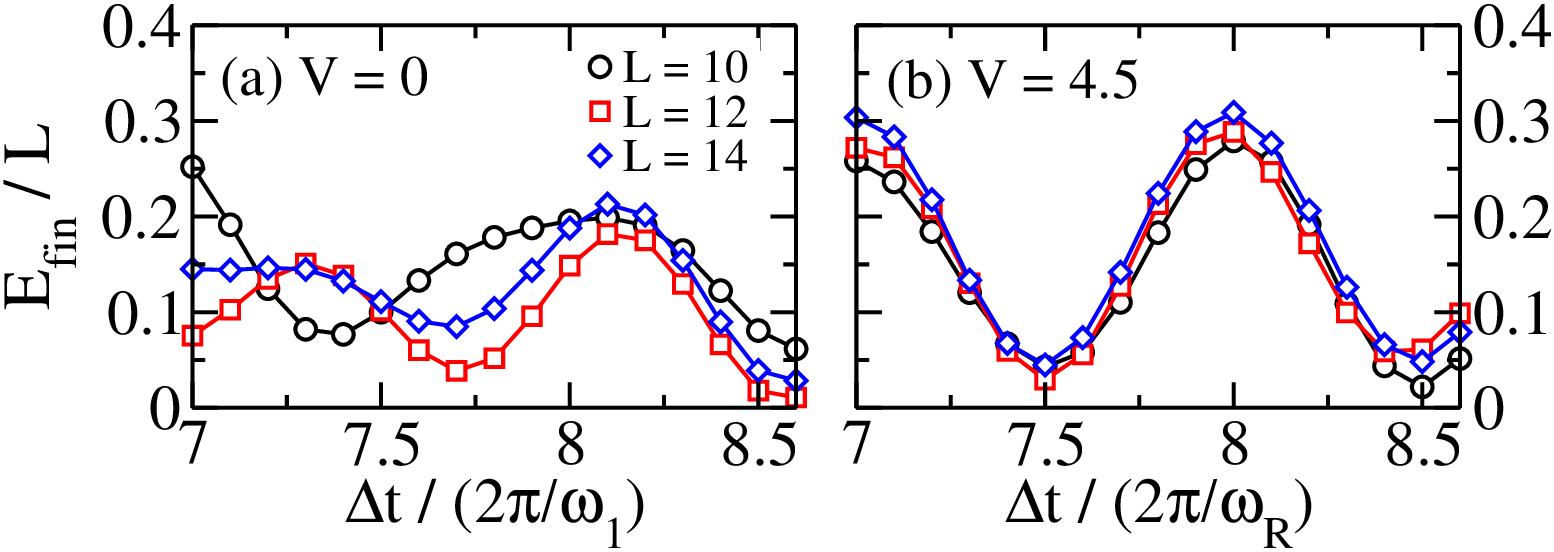}
\caption{(Color online) The final energy per site after the action of
  the second pulse, $E_{\text{fin}}/L$, as a function of $\Delta t$
  for various lattice sizes $L=10,\,12$ and $14$. (a) for $V=0$; (b)
  $V=4.5$. $\Delta t$ is normalized by $2\pi/\omega_1$ and
  $2\pi/\omega_{\text{R}}$, respectively. In all cases,
  $N_{\text{pt}}$ is fixed to $0.5$. The statements on how to specify
  other parameters, i.e., $\omega_{\text{pump}}$, $t_d$, and $A_0$,
  can be found in the main text.}
\label{fig:4}
\end{figure}

Figure~\ref{fig:4} shows the final energy per site $E_{\text{fin}}/L$
as a function of the normalized temporal interval $\tilde{\Delta}t$,
for various lattice sizes $L=10,\,12$ and $14$. In Fig.~\ref{fig:4}(a)
we present results for $V=0$, and in (b) for $V=4.5$, with
$\tilde{\Delta}t:= \Delta t/\left(2\pi/\omega_1\right)$ and $\Delta
t/\left(2\pi/\omega_{\text{R}}\right)$, respectively. The results for
$\tilde{\Delta}t$ ranged from $7$ to $8.6$ are presented. In clear
contrast to the almost perfect periodicity at $V=4.5$
(Fig.~\ref{fig:4}(b), where peaks (valleys) of $E_{\text{fin}}$ are
found independently of $L$ at the integer (half-integer) values of
$\tilde{\Delta}t$), Fig.~\ref{fig:4}(a) for $V=0$ shows a different
picture. Results for different $L=10$ through $14$ are consistent with
the admixture of higher harmonics and display large finite-size
effects. At $\tilde{\Delta}t=7$, large differences for different $L$
are found. After $\tilde{\Delta}t=8$, the curves show more consistent
results. However, with increase of $\tilde{\Delta}t$, results for
$E_{\text{fin}}$ start showing larger discrepancies again, e.g.,
around $\tilde{\Delta}t=11$ (not shown).

We suggest that there could be two reasons responsible for the
irregularities observed in Fig.~\ref{fig:4}(a). One is the mixing of
absorptions at different frequencies. Here, according to our setting,
at least two resonance frequencies are in the reach of the pumping
pulse. We can expect that with the increase of the system size
separated peaks in the optical spectrum gradually form a
continuum. Therefore, for a given $t_d$, the distribution of
optically-allowed states become continuous within the reach of pumping
frequencies, which in turn gradually diminishes the excited-deexcited
oscillations that appear in small-size calculations. Another reason
might be the decay of charge carrier pairs, due to the lack of the
protection of bound states. Let us suppose that a doublon-hole pair is
created, separately by a single lattice site. Subsequently, hopping of
the doublon (or the hole) to the neighboring sites leads to an
incoherent decay of the excitation into the doublon-hole
continuum~\cite{Hassler:2009be}. Compared with Fig.~\ref{fig:4}(b), we
speculate that it can be the main underlying mechanism for the
disparity between adjacent cycles found in Fig.~\ref{fig:4}(a).


In summary, we propose an optical interferometric experiment to detect
Rabi-like oscillations in a strongly correlated system, i.e., the
extended Hubbard model. Depending on the relative phase, it is
possible to observe the system being excited and then deexcited (or
enhanced further) by two consecutive laser pulses. A comparative study
of the Hubbard model without nearest neighbor interactions highlights
the important role played by the excitonic bound states in this
ultrafast dynamics. The key finding of this letter is that when a
precisely selected electric pulse in a correlated system triggers the
excitation of a many-body state that has no efficient decay path, a
quantum interference measurement may be realized within the
experimentally accessible time scale. Suggested experiments may lay
course for further research of coherent control and manipulations on
many-body systems.

\acknowledgments
  This work was also supported by the Strategic Programs for
  Innovative Research (SPIRE), the Computational Materials Science
  Initiative (CMSI), the global COE program ‘‘Next Generation Physics,
  Spun from Universality and Emergence’’ from MEXT, the Yukawa
  International Program for Quark-Hadron Sciences at YITP, Kyoto
  University, and SLO-Japan collaboration project from ARRS and
  JSPS. T.T. acknowledges support by the Grantin-Aid for Scientiﬁc
  Research (Grant No. 22340097) from MEXT. J.B. acknowledges support
  by the P1-0044 of ARRS, and CINT user program, Los Alamos National
  Laboratory, NM USA. A part of numerical calculations was performed
  in the supercomputing facilities in YITP and ACCMS, Kyoto
  University, and ISSP in the University of Tokyo.

\bibliographystyle{eplbib}
\bibliography{twopulses_nourl}

\begin{thebibliography}{10}
\expandafter\ifx\csname url\endcsname\relax\def\url#1{\texttt{#1}}\fi

\bibitem{Heberle:1995}
\Name{Heberle A.~P., Baumberg J.~J. \and K\"ohler K.} \REVIEW{Phys. Rev.
  Lett.}{75}{1995}{2598}.

\bibitem{Marie:1997}
\Name{Marie X., Le~Jeune P., Amand T., Brousseau M., Barrau J., Paillard M.
  \and Planel R.} \REVIEW{Phys. Rev. Lett.}{79}{1997}{3222}.

\bibitem{Stievater:2001}
\Name{Stievater T.~H., Li X., Steel D.~G., Gammon D., Katzer D.~S., Park D.,
  Piermarocchi C. \and Sham L.~J.} \REVIEW{Phys. Rev. Lett.}{87}{2001}{133603}.

\bibitem{Kamada:2001il}
\Name{Kamada H., Gotoh H., Temmyo J., Takagahara T. \and Ando H.} \REVIEW{Phys.
  Rev. Lett.}{87}{2001}{246401}.

\bibitem{Htoon:2002ee}
\Name{Htoon H., Takagahara T., Kulik D., Baklenov O., Holmes A.~L. \and Shih
  C.~K.} \REVIEW{Phys. Rev. Lett.}{88}{2002}{087401}.

\bibitem{Dagotto:2005ip}
\Name{Dagotto E.} \REVIEW{Science}{309}{2005}{257}.

\bibitem{Orenstein:2012bz}
\Name{Orenstein J.} \REVIEW{Phys. Today}{65}{2012}{44}.

\bibitem{Yamashita:1999bt}
\Name{Yamashita M., Manabe T., Kawashima T., Okamoto H. \and Kitagawa H.}
  \REVIEW{Coord. Chem. Rev.}{190-192}{1999}{309}.

\bibitem{Ogasawara:2000er}
\Name{Ogasawara T., Ashida M., Motoyama N., Eisaki H., Uchida S., Tokura Y.,
  Ghosh H., Shukla A., Mazumdar S. \and Kuwata-Gonokami M.} \REVIEW{Phys. Rev.
  Lett.}{85}{2000}{2204}.

\bibitem{Iwai:2003cj}
\Name{Iwai S., Ono M., Maeda A., Matsuzaki H., Kishida H., Okamoto H. \and
  Tokura Y.} \REVIEW{Phys. Rev. Lett.}{91}{2003}{057401}.

\bibitem{Nagai:2002}
\Name{Nagai M. \and Kuwata-Gonokami M.} \REVIEW{J. Phys. Soc.
  Jpn.}{71}{2002}{2276}.

\bibitem{Okamoto:2000cv}
\Name{Okamoto H., Kishida H., Matsuzaki H., Manabe T., Yamashita M., Taguchi Y.
  \and Tokura Y.} \REVIEW{Nature}{405}{2000}{929}.

\bibitem{Maekawa:2004}
\Name{Maekawa S., Tohyama T., Barnes S.~E., Ishihara S., Koshibae W. \and
  Khaliulin G.} \Book{Physics of Transition Metal Oxides} Vol. 144 of
  \emph{Springer Ser. Solid-State Sci.} (Springer, Berlin) 2004 pp. 68--79.

\bibitem{Stephan:1996}
\Name{Stephan W. \and Penc K.} \REVIEW{Phys. Rev. B}{54}{1996}{R17269}.

\bibitem{Gebhard:1997}
\Name{Gebhard F., Born K., Scheidler M., Thomas P. \and Koch S.~W.}
  \REVIEW{Philos. Mag. B}{75}{1997}{47}.

\bibitem{Shuai:1997}
\Name{Shuai Z., Pati S.~K., Su W.~P., Br\'edas J.~L. \and Ramasesha S.}
  \REVIEW{Phys. Rev. B}{55}{1997}{15368}.

\bibitem{Jeckelmann:2003}
\Name{Jeckelmann E.} \REVIEW{Phys. Rev. B}{67}{2003}{075106}.

\bibitem{Matsueda:2004ht}
\Name{Matsueda H., Tohyama T. \and Maekawa S.} \REVIEW{Phys. Rev.
  B}{70}{2004}{033102}.

\bibitem{Okamoto:1990}
\Name{Okamoto H., Toriumi K., Mitani T. \and Yamashita M.} \REVIEW{Phys. Rev.
  B}{42}{1990}{10381}.

\bibitem{Nakamura:2000gk}
\Name{Nakamura M.} \REVIEW{Phys. Rev. B}{61}{2000}{16377}.

\bibitem{Ejima:2007go}
\Name{Ejima S. \and Nishimoto S.} \REVIEW{Phys. Rev. Lett.}{99}{2007}{216403}.

\bibitem{Matsueda:2010ue}
\Name{Matsueda H., Sota S., Tohyama T. \and Maekawa S.} \REVIEW{J. Phys. Soc.
  Jpn.}{81}{2012}{013701}.

\bibitem{Lu:2012es}
\Name{Lu H., Sota S., Matsueda H., Bon{\v c}a J. \and Tohyama T.} \REVIEW{Phys.
  Rev. Lett.}{109}{2012}{197401}.

\bibitem{DeFilippis:2012jk}
\Name{De~Filippis G., Cataudella V., Nowadnick E.~A., Devereaux T.~P.,
  Mishchenko A.~S. \and Nagaosa N.} \REVIEW{Phys. Rev.
  Lett.}{109}{2012}{176402}.

\bibitem{park:5870}
\Name{Park T.~J. \and Light J.~C.} \REVIEW{J. Chem. Phys.}{85}{1986}{5870}.

\bibitem{Prelovsek:20111p}
\Name{{Prelov\v{s}ek} P. \and {Bon\v{c}a} J.} \Book{{Ground State and Finite
  Temperature Lanczos Methods}} arXiv:1111.5931 [cond-mat.str-el] (Nov. 2011).

\bibitem{Rabi:1936}
\Name{Rabi I.~I.} \REVIEW{Phys. Rev.}{49}{1936}{324}.

\bibitem{Rabi:1937}
\Name{Rabi I.~I.} \REVIEW{Phys. Rev.}{51}{1937}{652}.

\bibitem{Jaynes:1963fa}
\Name{Jaynes E. \and Cummings F.} \REVIEW{Proc. IEEE}{51}{1963}{89}.

\bibitem{Hassler:2009be}
\Name{Hassler F. \and Huber S.~D.} \REVIEW{Phys. Rev. A}{79}{2009}{021607}.

\end{thebibliography}

\end{document}